%% file: main.tex
\documentclass[twocolumn]{aastex701}

\usepackage{graphicx}
\usepackage{bm}
\usepackage{amssymb}
\usepackage{amsmath}
\usepackage{units}
\usepackage{array}

\usepackage[T1]{fontenc}
\usepackage{aecompl}
\usepackage{calligra}
\usepackage{soul}
\DeclareMathAlphabet{\mathcalligra}{T1}{calligra}{m}{n}
\DeclareFontShape{T1}{calligra}{m}{n}{<->s*[2.2]callig15}{}

\newcommand{\Msun}[0]{M_\odot}
\newcommand{\Mdot}[0]{\dot M}
\newcommand{\mdot}[0]{\dot m}

\newcommand{\edd}[0]{{\rm Edd}}
\newcommand{\eff}[0]{{\rm eff}}

\newcommand{\Angstrom}{\text{\AA}}

\newcommand{\es}[0]{{\rm es}}
\newcommand{\abs}[0]{{\rm abs}}
\newcommand{\bal}[0]{{\rm Bal}}
\newcommand{\effred}[0]{{\rm eff,\, red}}
\newcommand{\effblue}[0]{{\rm eff,\, blue}}
\newcommand{\Beff}[0]{{\mathcal{B}_{\rm eff}}}

\usepackage{booktabs}

\usepackage[dvipsnames]{xcolor}
\usepackage{orcidlink}

\shorttitle{Binary Balmer Breaks}
\shortauthors{Tiede \& Haiman}
\graphicspath{{./}{Figures/}}

\newcommand{\PaperBackmatter}{%
\begin{acknowledgments}
This work was supported by the European Union’s Horizon 2023 research and innovation program under Marie Sklodowska-Curie grant agreement No.~101148364.
This work is additionally supported by ERC grant (Bright-BHBs, grant agreement no.101200084). 
Views and opinions expressed are however those of the authors only and do not necessarily reflect those of the European Union or the European Research Council Executive Agency. 
Neither the European Union nor the granting authority can be held responsible for them.
The authors note that OpenAI's GPT-5.6 Sol was used as an aid in refining the manuscript text, developing and debugging plotting scripts, and clarifying the presentation of calculations and concepts. 
The scientific ideas, analysis, interpretation, and conclusions presented in this work are those of the authors, who reviewed and verified any generated material.
\vspace{8pt}
\end{acknowledgments}
\par
\software{
  \texttt{numpy} \citep{numpy},
  \texttt{scipy} \citep{scipy},
  \texttt{matplotlib} \citep{matplotlib},
  \texttt{cmasher} \citep{cmasher}
}
\bibliography{refs}{}
\bibliographystyle{aasjournalv7}
}

\begin{document}

\title{Balmer breaks in accreting super-massive black hole binaries}

\correspondingauthor{Christopher Tiede}
\email{christopher.tiede@nbi.ku.dk}
\author[0000-0002-3820-2404]{Christopher Tiede}
\affiliation{Niels Bohr International Academy, Niels Bohr Institute, Blegdamsvej 17, 2100 Copenhagen, Denmark}
\email{christopher.tiede@nbi.ku.dk}
\author[0000-0003-3633-5403]{Zolt\'an Haiman}
\affiliation{Institute of Science and Technology Austria (ISTA), Am Campus 1, Klosterneuburg 3400, Austria}
\email{zoltan.haiman@ist.ac.at}

\begin{abstract}

We show that truncated circumbinary accretion disks around super-massive black hole binaries can produce Balmer breaks. 
For binaries interacting with surrounding disks in a truncated, non-accreting phase, gas accumulates at the cavity edge, and can create conditions conducive to generating spectral discontinuities from $n=2$ hydrogen bound-free absorption.
We develop a simple model for predicting the presence of a Balmer edge and estimating the break strength, and we find that the strongest discontinuities are expected for binaries with periods of months to years and total masses $\gtrsim10^6\,\Msun$.
These systems share considerable overlap with those binaries anticipated to appear in upcoming time-domain surveys---like LSST and the Roman HLTDS---as periodically variable quasars.
We estimate that for typical feeding rates of a few tenths of their Eddington rate, a few-to-10\% of periodic binaries at $z\lesssim1$ should show prominent Balmer breaks, and that this fraction increases for the more massive black holes that can be discovered at higher redshift. 
We suggest that black hole binaries that truncate their accretion flows will appear as periodically variable sources, and will additionally be intrinsically X-ray weak, deficient in ionizing emission, and in favorable cases, produce notable Balmer edges in their spectra.
This combination provides a complementary observational test for massive black hole binary candidates.

\end{abstract}
%
%
\input{paper.tex}
%
\end{document}

%% file: paper.tex
%
\section{Introduction} 
\label{sec:introduction}

Spectral discontinuities from atomic bound-free transitions provide important diagnostics of the physical conditions of emitting material in astrophysical sources. 
The most prominent examples are the Lyman and Balmer edges of atomic hydrogen, whose appearance and strength depend sensitively on the temperature and ionization state of the emitting material. 
The Balmer break is particularly informative because substantial absorption from the \(n=2\) level of neutral hydrogen is produced only over a relatively narrow range of thermodynamic conditions. 
Balmer discontinuities are typically prominent in the spectra of A-type stars and, in integrated stellar populations, provide a useful tracer of intermediate-age stars \citep{BruzualCharlot:2003}. 
As a result, the strength of the Balmer break is commonly used to constrain the ages of galaxies and stellar clusters, recent star-formation histories, and redshifts \citep[e.g.,][]{Kauffmann:2003, Kriek:2006, RobertsBorsani:2024, Kuruvanthodi:2026}. 

Strong hydrogen bound-free discontinuities are not generally observed in the continua of AGN \citep{Antonucci:1989, Stevans:2014}, despite predictions from classical accretion-disk atmosphere models \citep[e.g.,][]{KoratkarBlaes:1999}. 
In particular, models for computing accretion disk emission by summing stellar spectra \citep[e.g.,][]{KolykhalovSunyaev:1984} or constructing vertical atmosphere profiles \citep[e.g.,][]{LaorNetzer:1989,Ross:1992,SincellKrolik:1998,BlaesAgol:1996,GenerozovHaiman:2014} can produce prominent hydrogen bound-free edges---though non-LTE effects, Comptonization, relativistic corrections, and more detailed treatments of the vertical disk structure can weaken or suppress these features \citep{ShieldsColeman:1994,Hubney:1997,Hubney:1998,Hubeny:2000,Hubeny2001,Blaes:2001}.
The general absence of prominent Lyman edges in observed AGN has become known as the ``Lyman edge problem'', though Balmer breaks have been observed in some sources \citep{Kishimoto:2003, Kishimoto:2004}.

Recently, in detailed radiation magnetohydrodynamics simulations, \citet{KaulJiangBlaes:2026} argued that such edges are suppressed in observed systems because of the low gas densities (and vertically inverted dissipation profiles) in magnetically supported disks.
However, in similar radiative transfer calculations, \citet{LiuQuataert:2025} demonstrated that AGN disks can produce prominent Balmer breaks, provided they are truncated at a radius where the disk effective temperature is qualitatively similar to the surface temperatures of typical A-type stars (and for even cooler temperatures in sufficiently low density atmospheres); though they comment that there is not an obvious reason for disks to truncate at such specific radii.

Hydrodynamical simulations of thin-disk accretion onto (super-) massive black hole binaries, though, have recently demonstrated that the rotating binary potential can halt circumbinary accretion flows on scales of the binary semi-major axis \citep{Ragusa+2016, Tiede:2025}; but that tidal and hydrodynamic heating of the inner-disk (i.e., at the edge of the central cavity) can keep these systems detectable in upcoming optical and infrared time-domain surveys \citep{Cocchiararo:2024, Tiede:2026}.
For binary configurations most relevant to identifying periodically variable sources in such surveys \citep[see, e.g.,][]{bogdanovicLRR+2022, Charisi:MMA:2022, DOrazioCharisi:Review:2023}, disk truncation occurs beyond the near-horizon scales where the standard ultraviolet, big-blue bump is generated.
This generally suppresses the ionizing continuum and makes an observable Lyman edge unlikely, but for sources that remain sufficiently luminous at optical wavelengths, it is possible that accreting (super-) massive black hole binaries with truncated accretion disks may present with notable Balmer edges in their spectra.

In this work, we estimate which accreting binary systems---assuming they truncate their accretion flow---would do so at disk radii that would result in a prominent Balmer break in the associated optical spectrum.
To accomplish this, we detail in Section~\ref{sec:disk-model} our disk modeling, and in Section~\ref{sec:balmer-breaks} our heuristic estimator for when Balmer breaks occur.
In Section~\ref{sec:binary-breaks} we apply these to estimate which combinations of binary mass, period, and feeding rate result in Balmer breaks and compare this with those binaries expected to be found in upcoming time-domain surveys like the Large Survey of Space and Time (LSST) and the Roman Space Telescope's High Latitude time-domain Survey (HLTDS).
We discuss or findings in the broader context of electromagnetic searches for (super-) massive black hole binaries and conclude in Section~\ref{sec:discussion}.

%
\section{Truncated disk model} 
\label{sec:disk-model}

We consider a circular binary of total mass $M_b$, semi-major axis $a_b$, and period $P_b$.
We assume that the binary is surrounded by a thin, Keplerian accretion flow with orbital frequency $\Omega = \sqrt{GM_b / r^3}$, but that mass transfer from the disk, across the circumbinary cavity, and onto the binary components is inefficient.
\citet{Tiede:2025} demonstrated that this is the case for geometrically thin accretion flows with disk aspect ratios $\lesssim 0.2$ \citep[see also, e.g.,][]{Ragusa+2016, Tiede2020, Heath+Nixon2020, Dittmann2022, Tiede:2026}.
For a system fed from large scales at a constant rate $\Mdot_\infty$, this means that the accretion rate onto the binary is $\Mdot_b \ll \Mdot_\infty$, such that material gradually accumulates against the inner-edge of circumbinary disk at $r_{\rm in}$.
We refer to this as the \emph{non-accreting limit}.
We detail the core elements of this picture and how it alters standard disk modeling below.

Far from the binary, the disk surface density is that of a steady-state disk with no boundary term,
\begin{align}
    \Sigma_0(r) = \frac{\Mdot_\infty}{3\pi \nu(r)} 
\end{align}
where $\nu(r)$ is the kinematic viscosity. 
We refer to this as the \emph{self-similar} solution.
The inner-region of the CBD is tidally coupled to the binary and receives angular momentum at a rate $\dot J = \ell_0 \dot M_b \sqrt{GM_b a_b}$, where the torque parameter $\ell_0$ is set by the specific hydrodynamics of mass transfer at the CBD inner edge.
As a result, the inner CBD surface density is modified away from the self-similar solutions as
\begin{align}
    \Sigma(r) = \frac{\Mdot_b}{3\pi \nu(r)} f(r) \ ,
 \label{eq:sigmaf}
\end{align}
with a boundary term $f \equiv \left( 1 + \ell_0 \sqrt{{a_b} / {r}} \right)$ whose significance is determined by the magnitude of $\ell_0$.\footnote{Note the sign change in $\ell_0$ as defined in \citet{Tiede:2025, Tiede:2026}, for convenience of interpretation.}
Following Appendix C in \citet{Tiede:2026} and assuming a fixed value for $\ell_0$, the disk is meaningfully modified away from the self-similar solution out to the viscous radius, $r_\nu$ \citep{Ivanov99, Rafikov2013, Rafikov2016}.
Interior to this radius, deviations of the inner-disk properties can be found by noting that
\begin{align}
    \Mdot_b = \Mdot_\infty / f(r_\nu) \ . 
\end{align}
For example, the associated density pile-up in the inner disk is
\begin{align}
    \Sigma(r_{\rm in}) = \chi \, \Sigma_0(r_{\rm in}) \ , 
\end{align}
where $\chi \equiv f(r_{\rm in}) / f(r_\nu)$.

We assume that the disk is in vertical hydrostatic and local thermodynamic equilibrium with characteristic scale-height $H$, such that the typical density is $\rho = \Sigma / (2H)$ and that the viscosity is described by a standard $\alpha$-prescription with
\begin{align}
    \nu(r) = \alpha \frac{P}{\rho} \frac{\beta^b}{\Omega} \ .
\end{align}
Here, $P = P_{\rm gas} + P_{\rm rad}$ is the sum of the thermal and radiation pressures respectively, $\beta = P_{\rm gas} / P$,  $\alpha$ is a dimensionless constant, and $b$ is a Boolean which sets if $\nu$ is proportional to only thermal pressure ($b=1$; $\beta$-disk) or the total pressure ($b=0$; $\alpha$-disk).
The effective emission temperature of the disk is given by
\begin{align}
    \sigma T_\eff^4 = \frac{3 \Mdot_\infty}{8\pi} \Omega^2 \chi \ ,
 \label{eq:teff}
\end{align}
for Stefan-Boltzmann constant $\sigma$,
and we adopt the solutions of \citet{Goodman:2003} (see also, \citealt{GoodmanTan:2004}), modified to accommodate the inner pile-up, 
\begin{align}
    \Sigma(r) &=
        \frac{2^{24/5}}{27\pi^{3/5}}
        \sigma^{1/5}
        \left( \mathcal{R} / \mu \right)^{-4/5} \\
        &\times \alpha^{-4/5} \kappa^{-1/5} \dot M_\infty^{3/5} \Omega^{2/5} \beta^{-4(b-1)/5} \chi^{3/5}, \nonumber \\
    H(r) &= \frac{3 \kappa \dot M_\infty}{8\pi c(1-\beta)} \chi .
 \label{eq:H_disk}
\end{align}
where $\kappa$ is the Rosseland mean opacity in the vertical energy diffusion equation, $\mu$ is the mean molecular weight, and $\mathcal{R} = k_B / m_H$ is the universal gas constant for hydrogen mass $m_H$ and Boltzmann constant $k_B$.
The disk model then separates into a piecewise solution with three regions: (1) the inner-most region where $P = P_{\rm rad}$ and $\kappa = \kappa_{\rm es}$, the electron scattering opacity; (2) the middle region where $P = P_{\rm gas}$ and $\kappa = \kappa_{\rm es}$; and (3) the outer region where $P = P_{\rm gas}$ and $\kappa = \kappa_{\rm abs}$, the mean absorptive opacity in the free-free regime (\citealt{SS1973}; see also, \citealt{HKM09}).
Of particular interest, in the radiation-pressure dominated region with $b=0$, the midplane density can be written directly in terms of the effective temperature as
\begin{align}
 \nonumber
    \rho_{\rm I} \simeq 2.1\times 10^{-9} \left(\mdot_\infty \chi\right)^{-3/2} \alpha_{0.1}^{-1}\, M_7^{-3/2}\, T_{8000}^{-2}\ {\rm g\,cm^{-3}} \ ,
\end{align}
where we've defined $\alpha_{0.1} \equiv \alpha / 0.1$, $M_7 \equiv M_b / (10^7 \Msun)$, $T_{8000} \equiv T_\eff / (8000\unit{K})$, and $\mdot_\infty \equiv \Mdot_\infty / \Mdot_\edd$ with $\Mdot_\edd(M_b)$ the Eddington accretion rate with radiative efficiency $\eta = 10\%$.

We can estimate typical values of $\chi$ by noting that, formally, $r_\nu$ grows slowly in time,\footnote{$r_\nu$ grows until it equals $\ell_0^2 a_b$ (as does $\chi$ as $f(r_\nu) \rightarrow 1$), but in the non-accreting limit, $\ell_0$ is often so large that achieving this steady-state exceeds the binary lifetime \citep{Tiede:2025, Tiede:2026}.}
such that in the non-accreting limit,
\begin{align}
    \Mdot_b(t) \simeq \frac{\Mdot_\infty}{\ell_0} 
    \left( 
        \frac{t}{t_\nu}
    \right)^{\lambda}
\end{align}
where $\lambda = (4 - 2n)^{-1}$ for power-law viscosities of the form $\nu \propto r^n$, and 
\begin{align}
    t_\nu = \frac{2}{3} \frac{a_b^2}{\nu(a_b)}
\end{align}
is the viscous time scale.
Thus, for $r_{\rm in} = \xi a_b$, the associated inner-density pile-up grows similarly,
\begin{align}
    \chi(t) \simeq \frac{1}{\sqrt{\xi}} 
    \left( 
        \frac{t}{t_\nu}
    \right)^{\lambda} \ .
 \label{eq:chi-t}
\end{align}
The amount of mass the system can accumulate at each separation, though, is bounded by the local residence time,
\begin{align}
    \left| \frac{\dot a_b}{a_b} \right|^{-1} 
    = \frac{M_b}{8 (\ell_0 + 3/8) \Mdot_b} 
    \simeq \frac{t_M}{8 \sqrt{\xi}} \chi^{-1} \ ,
 \label{eq:tres}
\end{align}
evaluated for a circular, equal mass binary,
where $t_M \equiv M_b / \Mdot_\infty \simeq 4.5 \times 10^7\,(\eta/0.1) \mdot_\infty^{-1} $ is the typical mass doubling time.
Thus, the disk will generally accumulate a pile-up of
\begin{align}
    \chi \simeq \frac{1}{\sqrt{\xi}}
    \left(
        \frac{t_M}{8 t_\nu}
    \right)^{\lambda / (1 + \lambda)} \ .
\end{align}
Specifically, for the radiation pressure dominated disk regions (and $b=0 $), $n = -3/2$ and $\lambda = 1/7$ giving a typical value,
\begin{align}
    \chi \simeq 2.6\ \mdot_\infty^{1/8} \, \alpha_{0.1}^{1/8} \, M_7^{1/6} \, \big( P_b /\unit{month} \big)^{-7/24} \ ,
 \label{eq:chi-estimate}
\end{align}
where we've taken $\xi = 3$.\footnote{At long binary periods, $t_\nu$ is very long, and $\chi$ can formally fall below unity. This simply means the disk does not have time to meaningfully deviate from the self-similar solution and one ought to take $\max(1, \chi)$.}

%
\begin{figure*}[t!]
    \centering
    \includegraphics{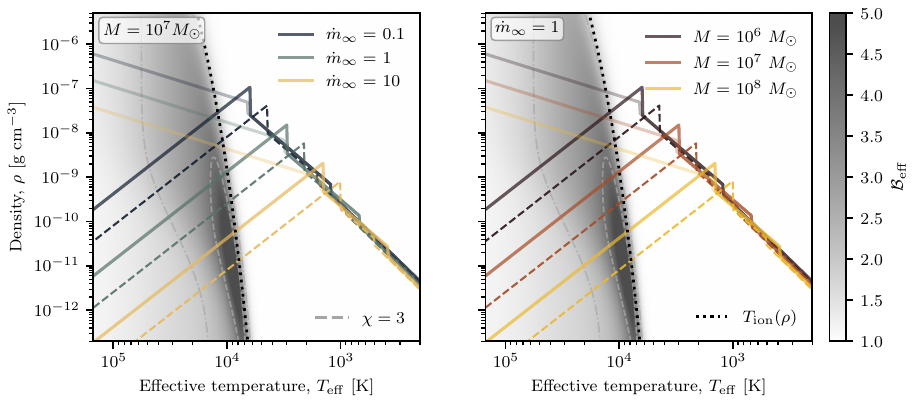}
    \caption{
    The effective opacity jump $\Beff$ (Eq.~\ref{eq:beff}) and disk density tracks (Section~\ref{sec:disk-model}) as a function of the effective temperature.
    The grey-scale regions show where Balmer breaks can become prominent ($\Beff > 1$), and the light-grey dot-dashed and dashed contours indicate $\Beff = \{2,\,4\}$, respectively.
    The black dotted curve shows the ionization temperature, $T_{\rm ioin}(\rho)$.
    The solid lines show the full disk solution for an $\alpha$ disk ($b=0$), while the transparent, monotonic lines show the solutions for a $\beta$ disk ($b=1$).
    The dashed lines illustrate $\alpha$-disk solutions with an inner-disk pile-up, $\chi = 3$.
    }
    \label{fig:beff}
\end{figure*}
%

%
\section{Balmer Break Estimation}
\label{sec:balmer-breaks}

Assuming hydrogen-dominated gas, two properties must generally be met in order for the emergent emission to posses a Balmer break: (1) the temperature must be high enough to populate the principal quantum number $n=2$ shell (L-shell) of hydrogen, but (2) it must not be so high as to fully ionize the gas.
Therefore, at a given gas density $\rho$, a Balmer break can only emerge when the gas temperature is very near the ionization temperature, $T_{\rm ion}(\rho)$.
Assuming that the system is in local thermodynamic equilibrium with ionization fraction $x$, the ionization temperature can be estimated by solving the Saha equation,
\begin{align}
    \frac{x^2}{1 - x} = 
    \left( \frac{2\pi m_e k_B T}{h^2} \right)^{3/2} 
    \frac{e^{-E_0 / k_b T}}{n_H} \ ,
 \label{eq:saha}
\end{align}
where $n_H = X_H \rho  / m_H$ is the hydrogen number density with $X_H = 0.75$, $E_0 = 13.6\,\unit{eV}$ is the ground state ionization energy,
$h$ is the Planck constant, and $m_e$ is the electron mass.

Conversely, solving Eq.~\ref{eq:saha} at a given temperature for the neutral hydrogen fraction, $1 - x$, allows one to estimate the number density of $n=2$ excited hydrogen through a Boltzmann equation,
\begin{align}
    n_2 = 4 (1 - x) n_H e^{- E_{21} / k_b T}
\end{align}
with $E_{\rm 21} = 10.2\,\unit{eV}$.
One can at this point, of course, appreciate how at sufficiently low temperatures $e^{-1/T} \rightarrow 0$ and at sufficiently high $T$, $x \rightarrow 1$, leading to no Balmer bound-free absorption.

Near the ionization temperature, we estimate the strength of the Balmer break by computing an effective opacity just red-ward and just blue-ward of the Balmer limit, $\nu_\bal = c / (3646 \Angstrom)$. 
Just red-ward of the limit, we estimate this opacity as
\begin{align}
    \kappa_\effred = [\kappa_\abs\,(\kappa_\abs + \kappa_\es)]^{1/2}
\end{align}
while just blue-wards we take
\begin{align}
    \kappa_\effblue = [(\kappa_\abs + \kappa_\bal)\,(\kappa_\abs + \kappa_\bal + \kappa_\es)]^{1/2} \ ,
\end{align}
%
where
\begin{align}
    \kappa_\bal = \frac{n_2 \sigma_\bal}{\rho} s_\bal(T)
\end{align}
is the approximate bound-free opacity at the Balmer edge.
The $n=2$ photo-ionization cross section is $\sigma_\bal = 1.6 \times 10^{-17} \,\unit{cm}^2$, and
$s_\bal(T) = 1 - e^{-h\nu_\bal / k_B T}$ corrects for stimulated emission at high temperatures.
We then estimate those regions likely to produce a Balmer break by computing the effective opacity jump
\begin{align}
    \Beff \equiv \frac{\kappa_\effblue}{\kappa_\effred}
 \label{eq:beff}
\end{align}
where values $\Beff > 1$ correspond to density-temperature conditions where $n=2$ bound-free absorption can generate a Balmer discontinuity.
Because of the reasoning at the beginning of this section, $\Beff$ ought to approach unity at both low and high-temperatures, and to only take larger values near $T_{\rm ion}$.
To accomplish this, we approximate the absorptive continuum phenomenologically as the sum of Kramer's' average opacity
plus the monochromatic free-free opacity near the Balmer edge \citep{RybickiLightman:RPA},
\begin{align}
    \kappa_\abs = \kappa_{\rm kr} + \kappa_{\rm ff, \nu_\bal}
\end{align}
with 
\begin{align}
    \kappa_{\rm kr} \simeq 5 \times 10^{24} \rho T^{-7/2} \, \unit{cm}^2 \unit{g}^{-1}
\end{align}
and
\begin{align}
    \kappa_{\rm ff,\nu_\bal} = \frac{4e^6}{3m_e h c}
        &\left( \frac{2\pi}{3m_e k_B T} \right)^{\! 1/2}
        \frac{(x n_H)^2}{\rho\,\nu_\bal^3} s_\bal(T)
 \label{eq:kappa_nub}
\end{align}
where $e$ is the elementary charge, and we've taken the Gaunt factor to be unity.
We include the monochromatic free-free opacity because at high temperatures, $\kappa_\bal / \kappa_{\rm kr} \propto T$, and $\kappa_{\rm ff, \nu_\bal}$ will more appropriately quantify the absorptive continuum in hot, ionized gas.
Thus, $\kappa_\abs$ may slightly overestimate the absorptive opacity near the Balmer edge, but we emphasize that $\mathcal{B}_{\rm eff}$ is a diagnostic rather than a direct prediction for the emergent flux ratio, and we use it to identify the regions of $(T,\rho)$ space favorable for producing a Balmer break.

We plot maps of $\Beff$ over the space of gas densities and effective temperatures as the shaded regions in Figure~\ref{fig:beff}.
White regions are consistent with $\Beff \sim 1$ and darker regions are those where Balmer breaks become possible and potentially prominent.
The dotted black line illustrates the ionization temperature as a function of $\rho$.
We over-plot in colored lines the disk solutions of Section~\ref{sec:disk-model}.
The left panel shows solutions for central (binary) masses $M = 10^7 \Msun$ at varying Eddington fraction, and the right panel solutions at fixed $f_\edd = 1$ for varying mass.
The solid lines are for $\alpha$-disk solutions ($b=0$), the translucent lines for $\beta$-disk solutions ($b=1$), and the dashed lines for $\alpha$-disks with inner-edge pile-ups corresponding to $\chi=3$.
We see that, as noted, Balmer breaks occur in the near vicinity of $T_{\rm ion}$ and are most prominent at lower disk density corresponding primarily to $\alpha$-disks around more massive binaries and higher Eddington fractions.

\begin{figure}[t!]
    \centering
    \includegraphics{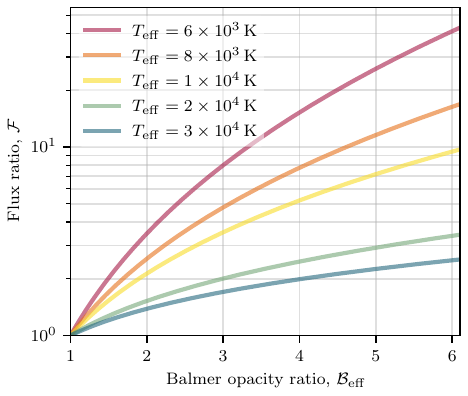}
    \caption{
    Estimates for the magnitude of flux ratio (Eq.~\ref{eq:flux}) across the Balmer edge---assuming constant opacities in the vertical atmosphere---as a function of the opacity jump, $\Beff$, for five values of $T_\eff$ where Balmer breaks are feasible (c.f., Figure~\ref{fig:beff}).
    Values of $\Beff \gtrsim 2-3$ generally correspond to significant Balmer breaks for these temperatures.
    }
    \label{fig:flux-jump}
\end{figure}

In our modeling, $\Beff > 1$ is a necessary condition for the presence of a Balmer discontinuity, but---absent some treatment of the vertical disk structure---it does not directly describe the magnitude of the actual flux difference across the Balmer edge.
We can, however, estimate the magnitude of the break by assuming a uniform density (and constant opacity), plane-parallel atmosphere with vertical coordinate $z$ measured down-ward from the disk surface.
Neglecting surface effects, the temperature profile follows $T^4(z) \propto z$ and the corresponding thermalization surface occurs at $z_\nu \approx (\rho \kappa_{\eff,\nu})^{-1}$.
As a result, the radiation temperature red-wards and blue-wards of the Balmer edge can be related by
\begin{align}
     T_\effblue^4  = T_\effred^4 / \Beff \ ,
\end{align}
and the flux discontinuity becomes,
\begin{align}
    \mathcal{F} (T_\eff, \Beff) \approx \frac{B_{\nu_\bal} [T_\eff]}{B_{\nu_\bal} [\Beff^{\! -1/4}\, T_\eff]}
 \label{eq:flux}
\end{align}
where $B_{\nu}$ is Planck's function, and we've tacitly taken $T_\effred$ as the reference temperature.
The Balmer frequency peaks at photosphere temperature $T_\bal \simeq h \nu_\bal / (2.8 k_B) = 1.4 \times 10^4\,\unit{K}$, so for temperatures lower than $T_\bal$, $\mathcal{F}$ is sampling the exponential Wien decay and the flux ratio will be highly sensitive to changes in radiation temperature.
At higher temperatures, the Balmer frequency moves towards the Rayleigh-Jeans tail where changes in photosphere temperature will produce a weaker flux discontinuity.
To visualize this, we plot values of $\mathcal{F}$ in Figure~\ref{fig:flux-jump} as a function of $\Beff$ for four values of $T_\eff$ where Balmer breaks might be observed.
We see that for relevant temperatures $\lesssim T_\bal$, as one moves blue-wards across the Balmer edge the flux drops by a factor of a few-to-ten once $\Beff \gtrsim 2$.
At higher temperatures, the contrast becomes expectedly much less pronounced.
We comment that in a more detailed treatment of the radiative transfer, these flux ratios are likely to be more modest, though the general trends should remain.
We conclude that the dark regions with $\Beff \gtrsim 2-3$ in Figure~\ref{fig:beff} represent systems most favorable for producing notable Balmer breaks.

%
%

%
\section{Binary Balmer Breaks}
\label{sec:binary-breaks}
%

%
\begin{figure*}[t!]
    \centering
    \includegraphics{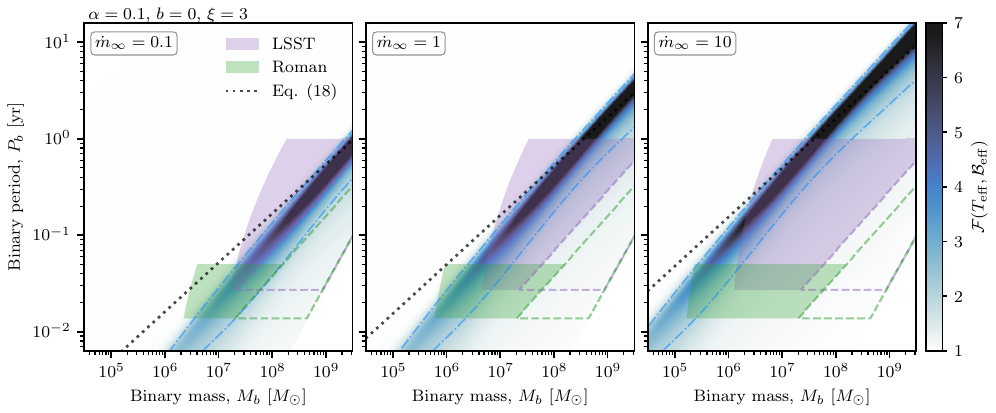}
    \caption{Balmer breaks in binary mass-period space for different outer feeding rates.
    Blue-to-black regions illustrate the magnitude of the estimated flux jump, and the $\mathcal{F} = 2$ contour is shown by the blue, dot-dashed line.
    The black-dotted line shows the estimate from Eq.~\eqref{eq:period-est}.
    The purple and green regions indicate estimates for the parameter space of binaries LSST and Roman can identify, respectively, as periodically variable sources at redshift $z=1$.
    The dashed regions show those binaries that change their frequency significantly over the survey lifetime (i.e., they chirp), but where this frequency evolution can enable detection with fewer observed cycles. 
    }
    \label{fig:binaries}
\end{figure*}
%

For massive black hole binaries surrounded by thin accretion disks in the non-accreting limit, material accumulates against the inner disk edge at $r_{\rm in} = \xi a_b$, such that Eq.~\eqref{eq:teff} becomes
\begin{align}
    T_\eff^4(r_{\rm in}) = \frac{3}{2}\frac{G M_b^2}{\sigma \kappa c \eta} \frac{\chi \mdot_\infty}{(\xi a_b)^3} \ .
\end{align}
A rough condition for the presence of Balmer breaks (from the relatively vertical $\Beff$ band in Fig.~\ref{fig:beff}) is $T_\eff \approx 8000K$, such that the rest-frame binary periods corresponding with truncated disks presenting with Balmer breaks can be estimated as
\begin{equation}
 \begin{aligned}
    P_b &\approx \left[ \frac{6 \pi^2 G M_b}{\sigma \kappa\, c\, \eta\, \xi^3} \frac{\chi \mdot_\infty}{T_\eff^4} \right]^{1/2} \\ 
    &\simeq 0.1 \, (\chi \mdot_\infty)^{1/2} M_7^{1/2} T_{8000}^{-2} \ \ \unit{yr} \, .
 \end{aligned}
 \label{eq:period-est}
\end{equation}
Thus, for our fiducial parameters, non-accreting binaries with periods of order months could posses Balmer breaks.

The actual manifestation of a Balmer break, though, is additionally sensitive to the gas density.
In Figure~\ref{fig:binaries}, from the disk modeling of Section~\ref{sec:disk-model} and the break strength estimates of Section~\ref{sec:balmer-breaks}, we plot the approximate flux jump across the Balmer frequency as a function of total binary mass and the (rest-frame) binary period for three values of the gas feeding rate, $\mdot_\infty = \{ 0.1, 1, 10 \}$.
Here, we input $\chi(\alpha, \mdot_\infty, M_b, P_b)$ according to Eq.~\ref{eq:chi-estimate}.
We plot our simple estimate from Eq.~\ref{eq:period-est} as the black-dotted line and see that it is a fair approximation for higher binary masses and feeding rates, but that it misrepresents the binary period in the low mass range.

We include in Figure~\ref{fig:binaries}, as purple and green shaded regions, estimates for those binaries expected to be detectable as periodically variable AGN in LSST and the Roman HLTDS. For LSST, we adopt a characteristic single-band cadence of $\sim 6$ days and a $\sim 10$ yr survey baseline, while for Roman we adopt the HLTDS cadence of $\sim 5$ days and a total baseline of $\sim 2$ yrs; the latter is expected to provide $\sim 6$ months of observations spread over this interval. 
The period limits are set by these cadences and survey durations, and we require that $>10$ orbital cycles be observed in order to confidently identify periodicity in the presence of stochastic AGN variability. 
The lower mass limits are estimated by truncating our standard, modified disk model at the inner-disk edge, computing the corresponding observed luminosity, and comparing this with the approximate single-visit survey sensitivities ($\sim 24\,\unit{mag}$ for LSST i-band and $\sim 27\,\unit{mag}$ for Roman F087). 
The upper mass limits are instead set by requiring that gravitational wave (GW) driven inspiral change the binary period by less than 10\% over the relevant survey baseline.
These sources would then vary with a single, generally consistent, period.
However, GW-driven frequency evolution can actually increase the confidence of a binary identification \citep{ChengCheng:chipring:2026}. 
We therefore indicate with dashed lines systems that may be detectable as chirping sources, where we have adopted a heuristic detection requirement to observe five cycles; though, we would not predict these sources to present with prominent Balmer breaks.

Figure~\ref{fig:binaries} shows considerable overlap between those binaries expected to appear in upcoming time-domain surveys and those that would posses Balmer breaks---especially for more modest feeding rates---but it is not readily clear how to map this to a number (or fraction) of sources these correspond to.
To estimate the fraction of observable sources that would have Balmer edges, we take the density of binaries at each mass, period, and redshift to scale as
\begin{align}
    \frac{dN}{d\log M_b\,d\log P_b\,dz}
    \propto \Phi(M_b,z) \,t_{\rm res}(P_b,M_b),
\end{align}
where $\Phi(M,z)$ is the comoving AGN mass function and $t_{\rm res}=P_b/|\dot P_b|$ is the binary residence time. 
We assume that the overall normalization is independent of binary mass and period so that it does not contribute to the fraction with Balmer breaks.
We adopt a simple power-law form for
\begin{align}
    \Phi(M,z)\propto M^{-\gamma},
\end{align}
as opposed to inferring a detailed active black hole mass function from the quasar luminosity function, because we believe this simple prescription makes the dependence on the underlying mass distribution transparent---and we find that our results are relatively insensitive to reasonable variations in $\gamma$.
We compute the residence time by adding the binary decay rates due to GW emission and gas driving.
The GW residence time scales as $t_{\rm GW}\propto M_b^{-5/3}P_b^{8/3}$, while Eqs.~\eqref{eq:tres} and~\eqref{eq:chi-estimate} give $t_{\rm gas}\propto\mdot_\infty^{-9/8}M_b^{-1/6}P_b^{7/24}$.
Of note, the gas-driven residence time depends much more weakly on both binary mass and orbital period than the GW-driven residence time, so that when gas driving dominates, low-mass and long-period binaries have much shorter residence times than they otherwise would from GW evolution alone.

%
\begin{figure}[t!]
    \centering
    \includegraphics{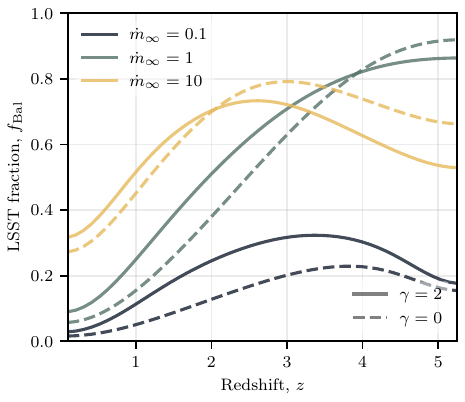}
    \caption{
    Estimate for the fraction of LSST sources that would posses Balmer breaks as a function of redshift for the same three outer feeding rates as in Figure~\ref{fig:binaries}.
    The dashed lines ($\gamma = 0$) correspond to a black hole mass function that is uniform in $\log M$, and the solid lines ($\gamma = 2$) to one that declines rapidly for larger BH masses.
    The fraction generally increases towards higher redshift as smaller-mass binaries become too faint for detection, but can turnover when the inner-disk becomes too hot.
    }
    \label{fig:fbal}
\end{figure}
%

We then restrict the binary population to the set of systems that satisfy the LSST detectability conditions in Figure~\ref{fig:binaries} and those that have break strengths $\mathcal{F} \geq 2$.
The former we denote with indicator function $S_{\rm LSST}$---which is unity for those systems that satisfy the constraints and zero otherwise---and the later with $S_\bal$.
Thus, the approximate fraction of detectable LSST binaries that would present with Balmer discontinuities is
\begin{align}
    f_\bal(z) = \frac{\int d\Gamma\, \Phi\, t_{\rm res} S_{\rm LSST} S_\bal}{\int d\Gamma\, \Phi\, t_{\rm res} S_{\rm LSST}} \ ,
\end{align}
where we have abbreviated $d\Gamma = d\log M d\log P$.
We plot this as a function of redshift in Figure~\ref{fig:fbal} for two contrasting values of \(\gamma=\{0,2\}\), corresponding to a mass distribution that is flat in $\log M$ and one that declines steeply toward larger BH masses.
We see that, because lower mass and longer period binaries are still generally more common (c.f., Figure~\ref{fig:binaries}), $f_\bal$ generally increases with growing feeding rate at all redshifts except the highest values of $\mdot_\infty$, where the detectable binaries at large-$z$ shift towards higher mass with inner-disk temperatures that become too hot to generate Balmer breaks.
For reference, we include a version of Figure~\ref{fig:binaries} with observing windows at $z=3$ in the Appendix.

At low redshifts, the Balmer fraction generally decays because the set of observable LSST binaries expands to the larger population of smaller mass black holes.
For high accretion rates, though, this can be somewhat counterbalanced by the associated decrease in $t_{\rm gas}$.
At high redshifts, the detectable population is dominated by the higher mass binaries where the Balmer break population tends to occupy a more populated region of the LSST accessible binary space (with the exception of the high accretion rate systems that become too hot).
Changing the power-law of the AGN mass function also does not meaningfully change these results, except for the lower $\mdot_\infty$ systems, because the detectable binaries shift strongly towards higher mass where there is significant overlap with the Balmer break binaries.
Thus, at low redshifts, we expect only a few-to-ten percent of LSST discovered periodic binaries to present with Balmer breaks---with the exception of systems being fed at many times their Eddington rate---but, that the more massive systems discovered at larger redshift become generally more likely to posses Balmer edges.

%
\section{Discussion and Conclusions}
\label{sec:discussion}

We have shown that standard, truncated circumbinary accretion disks can naturally occupy the narrow range of density and temperature conditions required to produce a prominent Balmer break. 
We additionally found that the growth of the inner-disk surface density and temperature (but decrease in volumetric density) from material piling up against the binary tidal barrier can additionally increase the set of sources with notable Balmer breaks.
We estimated that the binaries that will truncate their disks in regions favorable for Balmer edges have periods of order months-to-years, and that breaks are most likely observed in systems with total mass $\gtrsim 10^6 \Msun$---unless the feeding rate at large scales is meaningfully larger than the binary's Eddington rate where smaller mass systems with approximately week long periods become viable.

An important consequence, and the primary prediction, of these findings is that the portion of binary parameter space most favorable for generating prominent Balmer discontinuities overlaps significantly with those that are expected to be observed in upcoming optical and infrared time-domain surveys, LSST and the Roman HLTDS, respectively \citep[e.g.,][]{WittCharisi:2022,Haiman:RomanWP:2023,ChengCheng:LSST:2024,BertassiDotti:2026}.
The fraction of possible periodic LSST and Roman binaries that would appear with Balmer breaks is primarily sensitive to the outer feeding rate.
Assuming that binary systems generally feed at a few-times-ten percent of their Eddington rate, we estimate that at redshifts $z \lesssim 1$ only a few-to-10\% of periodic binary sources will have Balmer discontinuities; but that the more massive sources at higher redshift become increasingly more likely to present with a Balmer edge (e.g., Figure~\ref{fig:fbal}).

Thus, the general picture predicted for super-massive black hole binaries interacting with accretion flows in a truncated, non-accreting phase would be AGN-like systems that are periodically variable in the optical to near-infrared, but because of the disk truncation, these sources would be markedly less luminous at high photon energies, leading to a number of characteristics compared against typical AGN:
\begin{itemize}
    \item they would be intrinsically X-ray weak and would not appear in X-ray observations, 
    \item as a result, they would lack high ionization emission lines typically excited by the inner regions of the central engine, 
    \item they may have under-luminous or transient standard broad emission line regions, and 
    \item they may be accompanied by notable Balmer breaks in their optical spectra.
\end{itemize}
A Balmer discontinuity would therefore provide further evidence for the presence of a binary, as well as a complementary spectroscopic test of the truncated, weakly accreting binary model.
Conversely, the systematic absence of such features in a population of candidate binaries occupying the favorable regions of Figure~\ref{fig:binaries} could place constraints on long-lived non-accreting phases, the amount of material accumulated at the cavity edge, the vertical structure of the disk, and the applicability of $\alpha$-disk models to such systems.

It is also intriguing that several of these predicted properties overlap with those of the recently identified ``little red dots'' (LRDs) in JWST observations. 
LRDs are generally remarkably weak in X-rays, and a significant subset exhibit strong Balmer breaks and Balmer absorption features \citep[e.g.,][]{Kocevski:LRDs:2023, Matthee+2024-LRD, Yue:LRDs:2024, Akins:2024, Setton:2025}. 
These properties have proven difficult to reconcile within standard AGN interpretations, and some of these challenging features can be achieved in a binary scenario (see also, \citealt{Tiede:2025, Inayoshi:bLRDs:2025}).
However, LRDs seem to show little to no variability \citep{Furtak:2025,Secunda:LRDs:2026,LiuNaiduSecunda:2026} (however, \citealt{Lambrides:LRDvary:2026}) and are generally identified by their very prominent broad emission lines \citep{Greene:LRDs:2024,Rusakov:2026,Sneppen:2026}, which would make an accreting binary scenario difficult to justify.

Our estimates remain deliberately simple, and a proper determination of the presence and strength of Balmer breaks (or other spectral features) would require a more detailed treatment of both the vertical structure of circumbinary accretion disks in such non-accreting phases and a radiative transfer procedure that incorporates a more formal treatment of the disk opacities and non-LTE effects.
Our modeling has also assumed that binary non-accretion phases persist into radiation pressure dominated disk regimes, which has yet to be directly demonstrated in simulations (though a key component in these results is the magnitude of the disk aspect ratio, which has been observed to remain relatively small in recent radiation-hydrodynamics and radiation-magnetohydrodynamics simulations; \citealt{Cocchiararo:2025, Tiwari:2025}).
Similarly, these models have not treated the magnetic fields in the accretion flow, which might meaningfully alter the nature of the binary non-accreting phases as well as the structure of the vertical disk atmospheres \citep[e.g.,][]{Most:MHD:2024, KaulJiangBlaes:2026}.

Taken together, these results suggest that Balmer breaks could provide a useful spectroscopic complement to time-domain searches for compact super-massive black hole binaries. 
Systems showing periodic optical or near-infrared variability together with weak high-energy emission and a pronounced Balmer break would form a particularly compelling subset for follow-up. 
More generally, combining variability, broadband spectral shape, and detailed spectroscopy offers a direct and promising way to test whether compact binaries spend an appreciable fraction of their evolution in truncated, non-accreting phases and to possibly identifying super-massive black hole binary candidates with next-generation instruments.

\PaperBackmatter

\appendix
\section{Redshift dependence of LSST \& Roman sources}
\label{app:appendix}

To illustrate how the set of binaries expected to appear in LSST and the Roman HLTDS evolves with redshift, we re-create Figure~\ref{fig:binaries} with $z=3$ used for the observing regions.

%
\begin{figure*}[h]
    \centering
    \includegraphics{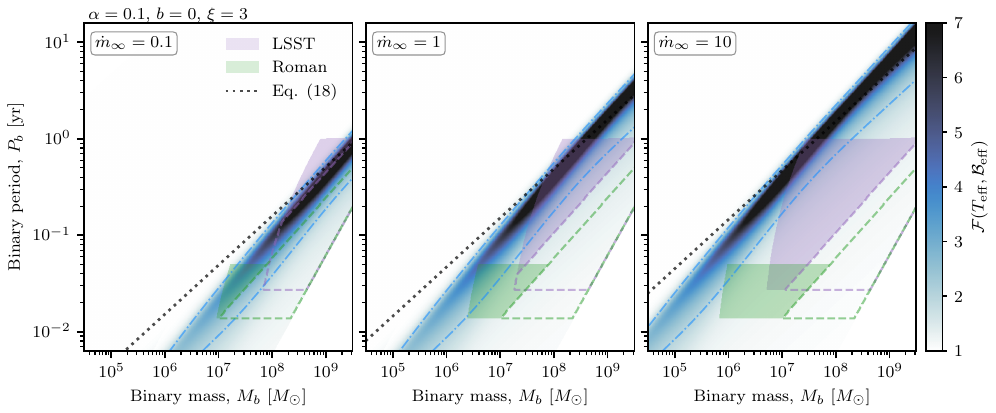}
    \caption{Same as Fig 3, but for sources at redshift $z=3$.
    }
\end{figure*}
%